# Data Normalization Strategies for EEG Deep Learning

Dung Truong[1], Arnaud Delorme[1,2]

*Abstract*— Normalization is a critical yet often overlooked component in the preprocessing pipeline for EEG deep learning applications. The rise of large-scale pretraining paradigms such as self-supervised learning (SSL) introduces a new set of tasks whose nature is substantially different from supervised training common in EEG Deep Learning applications. This raises new questions about optimal normalization strategies for the applicable task. In this study, we systematically evaluate the impact of normalization granularity—recording vs. window level—and scope—cross-channel vs. within-channel—on both supervised (age and gender prediction) and self-supervised (Contrastive Predictive Coding) tasks. Using high-density resting-state EEG from 2,836 subjects in the Healthy Brain Network dataset, we show that optimal normalization strategies differ significantly between training paradigms. Window-level within-channel normalization yields the best performance in supervised tasks, while minimal or cross-channel normalization at the window level is more effective for SSL. These results underscore the necessity of task-specific normalization choices and challenge the assumption that a universal normalization strategy can generalize across learning settings. Our findings provide practical insights for developing robust EEG deep learning pipelines as the field shifts toward large-scale, foundation model training.

*Keywords—Deep Learning, EEG, Normalization*

## I. Introduction

Deep Learning has been widely applied to EEG data to great success [1]. Much attention has been placed on instigating the ability of deep models to reduce the needs for EEG feature engineering [2, 3] and for signal processing to accommodate noisier data [4, 5]. However, an often glossed over aspect of feature exploration for EEG deep learning is normalization. Deep learning's success in vision and language has hinged on carefully chosen normalization schemes—BatchNorm, LayerNorm and their variants—showing that normalizing layer inputs is a core architectural decision, not just an implementation detail [6, 7]. For EEG deep learning applications, short-time windows are often extracted from each long-time EEG recording to boost the number of samples or to reduce dimensionality of input [1]. Each multi-channel EEG window becomes an input to the neural network. Current practices largely applied a one-size-fits-all recording-level, per-channel normalization, without questioning whether window-level or cross-vs-within-channel scaling might yield better representations.

In addition, with the increase in standardization and data sharing efforts [8-11], and the advances of deep learning in building foundation models [12], the field of EEG is moving towards large-scale training using more general training paradigms such as Self Supervised Learning [13-18]. It has not been investigated whether normalization approaches for the supervised tasks of prior EEG deep learning works would transfer well to the general learning paradigm across datasets without specific target labels.

In this paper, we present a comprehensive empirical study that examines how different normalization strategies affect performance in both supervised and self-supervised EEG deep learning tasks. Using the large-scale Healthy Brain Network (HBN) dataset comprising high-density resting-state EEG from 2,836 subjects [19, 20], we systematically evaluate the impact of normalization applied at different levels (recording vs. window) and scopes (cross-channel vs. within-channel). Supervised learning is assessed via age regression and gender classification, while SSL performance is measured using a Contrastive Predictive Coding (CPC) pretext task [21].

Our findings suggest that normalization should not be treated as a one-size-fits-all operation and must be carefully selected to match the learning objective. As the EEG research community moves toward large-scale, cross-task, and cross-domain learning with foundation models, our work provides critical insights into the design of normalization strategies that can support robust and scalable EEG deep learning.

## II. Methods

**Dataset.** The dataset originates from the Healthy Brain Network (HBN) project at the Child Mind Institute, a large-scale initiative aiming to build a community-based biobank of data from up to 10,000 children and adolescents (ages 5-21) with rich psychopathology assessments [19]. Participants underwent various tasks while multimodal brain imaging (MRI, fMRI, and EEG), eye-tracking, actigraphy, and voice and video data were collected. The complete dataset also encompasses behavioral, cognitive, and genetic information. The dataset has been reformatted into the standardized BIDS format for efficient sharing and processing, called HBN-EEG [20]. For this study, we utilized a subset of resting-state EEG data from 2,836 participants across all 11 public releases of HBN-EEG. During this 5-minute task, subjects viewed a fixation cross and were prompted to open or close their eyes at different times. High-density EEG data were recorded at 500 Hz with a 0.1-100 Hz bandpass in a sound-shielded room using a 128-channel EGI geodesic hydrogel system. Only the 6-minute resting-state data, consisting of alternating eyes-open and eyes-closed periods, was used.

**EEGDash archive.** EEGDash is a collaborative initiative between the Swartz Center for Computational Neuroscience (SCCN) at UC San Diego and Ben-Gurion University, supported by the National Science Foundation. It serves as a comprehensive data-sharing platform for electrophysiology data, aiming to facilitate machine learning (ML) and deep learning (DL) applications. The archive aggregates data from 25 laboratories, encompassing over 27,000 participants, including both healthy individuals and clinical populations. EEGDash provides data in PyTorch-compatible formats, integrating with the Braindecode library for preprocessing, thus streamlining the development of DL models [24]. The platform also offers a Python package, 'eegdash', available via PyPI, enabling users to query and download datasets efficiently. This study utilized EEGDash datasets ds005505 through ds005516 (release 1 to 11) of the HBN-EEG data.

---
[1] Swartz Center for Computational Neuroscience, Institute for Neural Computation, UC San Diego, La Jolla, CA, USA
[2] CerCo CNRS, Paul Sabatier University, Toulouse, France

**Data splitting.** We split the data into training and test sets using the natural splits of different releases. To have sufficient training data to avoid overfitting, we used a relatively small release, 6 (ds005510), as the test set. No validation set was sampled as we did not perform any hyperparameter search. The rest of the data was used as the training set. This splitting approach avoids data leakage as there is no subject overlap among the releases. In total, we have 2,567 subject recordings for training and 134 recordings for testing.

**Preprocessing.** Although DL may be applied to raw EEG data without any preprocessing, filtering can have a dramatic influence on signal quality and performance [3, 22]. As the dataset is identified to be heavily contaminated with line noise, we notch filter the 60Hz and its harmonic 120Hz line noise frequencies. Using Braindecode [24], we bandpass filtered the data between 0.1 Hz and 59 Hz and resampled the data to 250 Hz.

For supervised tasks, we segmented resting data periods into non-overlapping 2-s windows: each preprocessed 2-s epoch was used as a sample for our final dataset. We removed the 0-value Cz reference channel, and as a result, each sample in our dataset thus had dimension 128x500 (128 channels and 2(s) x 250(Hz) time points). No further preprocessing is done at the recording level or window level.

**Normalization strategies.** While standard normalization (0 mean and 1 standard deviation) and min-max normalization are two common methods for feature scaling in deep learning, we applied robust normalization to account for outliers in the EEG signal, including extremely large voltages due to noisy bursts of the data. Robust scaling uses the median and the interquartile range (the difference between the 75th quantile and 25th quantile) instead of mean and standard deviation as with standard scaler, hence less sensitive to outliers.

We tested two levels of normalization: recording level and window level. For each level, we compared three normalization schemes: no normalization, cross-channel, and within-channel normalization. Figure 1 depicts these different normalization levels and approaches. As their names indicate, the cross-channel approach uses statistics of the entire recording/window, whereas the within-channel normalization uses statistics of the channels individually. As a result, cross-channel normalization preserves the relative scaling relationship between channels' voltage, while within-channel normalization only preserves the temporal relationship of the timepoints within each channel.

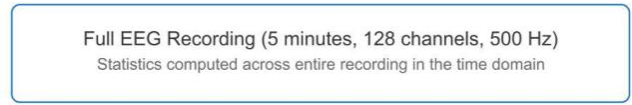
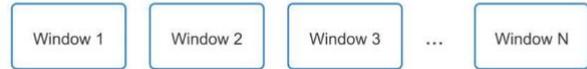
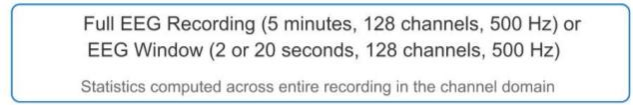
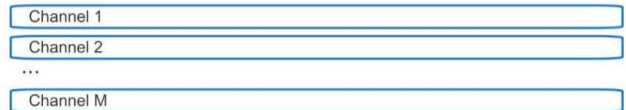

*Figure 1.* Illustration of normalization strategy taxonomies evaluated in the study. (A) Recording-level vs. window-level normalization: normalization statistics are computed either across the full EEG recording (global temporal context) or within each temporal window (local context). (B) All-channel vs. channel-level normalization: normalization statistics are computed across all channels jointly (preserving spatial scale relationships) or independently per channel.

**Supervised tasks.** We use two subject attribute prediction tasks, age and gender prediction, to evaluate the impact of normalization techniques on supervised training. These two metrics (age and gender) have been shown to be sufficiently predictable by raw resting state EEG data [2, 23]. For gender classification, we produced a subset of the dataset to contain a balanced male/female distribution, as in prior work [2]. All reports on gender classification henceforth are on balanced subsets of the data.

**Self-supervised task.** Self-Supervised Learning (SSL) is a deep learning paradigm that allows model training without human-provided labels [12]. SSL uses a *pretext task* that leverages the structure within the data itself to generate labels. Research on SSL for EEG data is still at its nascent stage, with very few works that show a pretext task that can be trained across datasets and produce generalizable embeddings that can be transferred to unseen tasks and datasets - a holy grail of SSL training. Among the pretext tasks that have been applied to EEG [14, 17, 31], Contrastive Predictive Coding (CPC) is a promising approach whose efficacy has been shown with other time-series data [21] as well as to EEG with good transfer learning performance [17]. For each EEG window sample, CPC uses an encoder to generate a sequence of vector embeddings representing the temporal dynamics of the input in the latent embedding space. It then randomly selects a portion of these embeddings to be replaced by a shared, learnable mask embedding. A contextualizer is applied to the output embeddings of the encoder, both masked and unmasked. For each masked instance, the contextualizer produces a guess, using information from other embeddings in the sequence, of the true unmasked version of the embeddings. This guess is then contrasted against a pool of distractors. The distractors are sampled either from within the sequence itself (of the same recording) or from different recordings. For CPC with same-recording distractors, we set the window size to be 20 seconds instead of 2 seconds for sufficient extraction of distractor samples. The goal of CPC training is to make the embedding of the predicted instance more similar to the true one, and less similar to the distractors [21]. We adapted the setup of [17] and set the mask rate to 0.1, mask span 1, and the number of distractors to 20. The rest of the parameters were kept the same.

**Neural network architectures.** For the supervised tasks, we employed the Deep4Net architecture that has been tested with similar tasks [23]. Deep4Net is a general convolutional architecture that combines temporal and spatial convolution, specifically designed for EEG data [24]. For supervised tasks, we employed the model's implementation in Braindecode as is [24], changing the number of outputs in the classification head according to the tasks (2 for binary gender classification and 1 for age regression). For the Contrastive Predictive Coding self-supervised task, we repurposed Deep4Net by removing the last classification module and not flattening the output of the convolutional layers to preserve the sequential dynamics of the input – CPC employs a contextualizer on top of the encoder. While previous work [17] used a Transformer architecture for the contextualizer [25], it is a complex architecture with potentially much different sensitivity to normalization. We thus chose a simpler model for the contextualizer, using a 3-layer unidirectional LSTM [26] for our experiments.

**Deep learning framework.** We structured our experiments using the EEG-SSL framework in [27], coupled with the infrastructure provided by PyTorch Lightning for easy experiment management and reproducibility [28]. Model architectures and preprocessing pipelines were implemented using PyTorch [29] and Braindecode [24].

**Evaluation metrics.** To compare the different normalization methods on different supervised and self-supervised tasks, we used the different assessment metrics that have been reported for each appropriate task. We used the mean absolute error for age regression and balanced accuracy for the gender classification at the sample level. For CPC, we use the cross-entropy loss of CPC for evaluation. In all cases, we report the best error/accuracy/loss value assessed on the test set.

**Experiment design.** We conducted a 3-by-3 experiment design where we compared the interactions between recording/window levels and three different normalization approaches as described above. For each combination, we trained 5 different models initialized with 5 different seeds and reported the means and standard deviations of the appropriate evaluation metrics across seeds. The batch size for supervised tasks was set at 64 and for CPC at 128. Models were trained using Adamax [30] with a learning rate of 0.002. Supervised experiments were run for 5 epochs, based on visual inspection indicating that the loss had stabilized. Self-supervised experiments were run for 10 epochs. All experiments were run on a single NVIDIA 4090-Ti GPU.

Our code is made publicly available at https://github.com/sccn/eeg-ssl.

### III. RESULTS

We observed patterns between training paradigms and between tasks of the same paradigm. We report our results in Tables 1-4. As shorthand, we will call no normalization approach *None,* cross-channel normalization *All,* and within-channel normalization *Channel* in our reports. For all tables, recording-level factors are reported in rows and window-level factors are reported in columns. The best-performing value of each task is highlighted with a gray background.

| Recording \Window | None | All | Channel |
|---|---|---|---|
| None | 1.40 ± 0.465 | 0.65 ± 0.183 | 3.04 ± 0.002 |
| All | 2.98 ± 0.114 | 2.95 ± 0.148 | NaN |
| Channel | 3.01 ± 0.006 | 3.01 ± 0.005 | 3.04 ± 0.002 |

Table 1. Mean and standard deviation of CPC loss with distractors sampled from the same recording.

| Recording \Window | None | All | Channel |
|---|---|---|---|
| None | 3.09 ± 0.001 | 3.84 ± 1.12 | 2.91 ± 0.19 |
| All | 3.05 ± 0.01 | NaN | NaN |
| Channel | 3.05 ± 0.01 | 3.06 ± 0.005 | 2.98 ± 0.11 |

Table 2. Mean and standard deviation of CPC loss with distractors sampled from different recordings.

| Recording \Window | None | All | Channel |
|---|---|---|---|
| None | 3.29 ± 0.14 | 1e+9 | 2.59 ± 0.44 |
| All | 3.17 ± 0.03 | NaN | NaN |
| Channel | 3.17 ± 0.04 | 3.15 ± 0.01 | 2.48 ± 0.36 |

Table 3. Mean absolute error (MAE) in years for age regression with mean and standard deviation. Lower values indicate better performance.

| Recording \Window | None | All | Channel |
|---|---|---|---|
| None | 0.50 ± 0.02 | 0.62 ± 0.04 | 0.83 ± 0.06 |
| All | 0.49 ± 0.02 | 0.50 ± 0.02 | 0.52 ± 0.00 |
| Channel | 0.48 ± 0.00 | 0.48 ± 0.00 | 0.85 ± 0.03 |

*Table 4. Balanced accuracy for gender classification with mean and standard deviation. Higher values indicate better performance.*

**CPC.** For CPC with same-recording distractor sampling (Table 1), no normalization surprisingly gave lower loss values than many other normalization permutations. The best result was obtained when no normalization was done in the recording level, but performed cross-channel at the window level (None-All). Within-channel normalization at any level performed poorly, suggesting it disrupts global structural signals needed for CPC. Other normalization combinations gave relatively similar performance, except for combining recording-level cross-channel normalization and window-level within-channel normalization (All-Channel) led to training collapse (NaNs).

For CPC with different-recording distractor sampling (Table 2), we observed failure mode in the All-All and All-Channel scenarios, rendering NaN losses, pointing to incompatibility of mixed normalization scopes for this task. Although the better results were obtained when normalization was done channel-wise at window-level (None-Channel and Channel-Channel), the differences were within the room of standard deviation, hence not significant.

**Age regression.** Channel-wise normalization has a clear impact on learning (Table 3). When there was no window normalization (Window-None column), normalizing the recording improved prediction. Window-level within-channel normalization after within-channel normalization at the recording level (Channel-Channel) gives the best performance for this task. We observed the same failure mode as with CPC with different-recording distractors, where All-Channel results in NaNs, confirming the incompatibility of mixed normalization scopes.

**Gender classification.** As with the age regression task, we observed that window-level within-channel normalization was crucial for achieving good prediction performance for gender classification (Table 4). No other normalization combination yielded useful learning for this task. This mirrors the age regression results, reaffirming the benefit of localized, channel-specific normalization for supervised classification tasks.

## IV. DISCUSSION

This study highlights the critical impact of normalization strategy on EEG deep learning performance, with distinct patterns emerging for supervised and self-supervised paradigms.

Our key findings are:

1. Window-level within-channel normalization significantly improves performance for two supervised EEG deep learning tasks. In contrast, recording-level normalization alone proved insufficient, and models trained without window-level normalization underperformed.
2. Self-supervised learning via Contrastive Predictive Coding with the same-recording distractor is more sensitive to local normalization, with cross-channel window-level normalization or no normalization outperforming other strategies.
3. Self-supervised learning via Contrastive Predictive Coding with distractors from across recordings exhibits similar failure modes as supervised tasks, showing a relationship between supervised and self-supervised tasks with respect to task structure.
4. A consistent failure mode across tasks occurred where recording-level cross-channel normalization combined with window-level within-channel normalization led to training collapse, yielding NaN losses.

Overall these findings show that across tasks, better results were obtained when normalization is done on the window level. When combining recording and window-level normalization, training starts to become unstable. The failure mode suggests that interaction likely disrupts internal consistency in feature scaling, impairing model training.

**Within vs. across recordings.** The nature of the learning tasks played a significant role in choosing appropriate normalization approaches. CPC with same-recording sampling inherently relies on sequential dynamics in each individual recording, whereas CPC with different-recording sampling and the supervised tasks might try to capture statistical signals that are consistent across recordings. This difference could contribute to the contrasting normalization outcomes observed. In tasks that require recording-level consistency, like same-recording CPC, preserving relative inter-channel relationships may be more important than reducing local variability, whereas for across-subject tasks, controlling for individual differences through localized normalization may be more beneficial. These distinctions highlight the value of aligning normalization strategies with both the task's temporal resolution and its underlying structure.

**Window size.** A consideration in interpreting the results is the difference in window size between the same-recording CPC and the other tasks. CPC with distractors sampled from the same recording was trained on 20-second windows, whereas other tasks used 2-second windows. While longer windows might be expected to support more stable normalization due to the increased amount of data per sample, within-channel normalization still performed poorly in this setting. We suspect that larger windows might lead to higher chances of containing extremely noisy EEG signals. It could also be that longer windows alone do not guarantee improved signal stability for all learning paradigms, particularly when the task depends on consistency among windows across recordings as discussed above.

**Dataset choice.** While this study relies exclusively on the HBN-EEG dataset, which may limit generalizability, the dataset itself includes a large and diverse cohort spanning a wide age range, multiple recording sessions, different technicians and hardware, and varying recording conditions. These factors introduce a degree of natural variability that partially mitigates concerns about overfitting to a narrow data distribution. Nevertheless, the findings are specific to resting-state, non-clinical EEG. Future studies should assess the reproducibility of these results across broader datasets and experimental contexts.

**Data cleaning.** Advanced EEG data cleaning methods such as Independent Component Analysis (ICA) [34] and the *clean_rawdata*() [35] pipeline were not used in this study, as the focus was specifically on normalization strategies rather than comprehensive artifact removal. ICA is a widely used blind source separation technique for identifying and eliminating non-neural artifacts like eye blinks and muscle activity from EEG data. The *clean_rawdata* function is an automated preprocessing pipeline that integrates steps such as bad channel detection, artifact subspace reconstruction (ASR) to improve EEG signal quality. While these techniques are not directly related to normalization, the use of advanced EEG rejection methods could enhance signal integrity and potentially interact with normalization approaches. Exploring this interaction represents a valuable avenue for future research. However, their exclusion is justified here: such methods have not been shown to significantly alter statistical outcomes [22], and comparable approaches like *autoreject* in MNE have been reported to reduce the performance of deep learning models [3].

**Model choice.** Though we expect our findings to generalize to other tasks of a similar nature, we recognize the difference in sensitivity to normalization for different neural network architectures. Particularly, Transformer is a much harder model to train and process local/global information much differently than LSTM. Thus, careful examination should be done when choosing normalization methods for other neural architectures.

**Self-supervised learning choice.** While our use of Contrastive Predictive Coding (CPC) provides a valuable and well-motivated framework for evaluating self-supervised learning on EEG data, there are some limitations worth noting. We evaluated model performance solely based on the CPC loss during pre-training, which, while informative, does not fully capture the downstream utility of the learned representations. Incorporating fine-tuning or transfer learning tasks would offer a more comprehensive assessment. To broaden the generalizability of these findings, future work should explore additional self-supervised approaches such as Relative Positioning [31], SimCLR [32], or various forms of autoencoders (e.g., variational, denoising) [33], which may exhibit different sensitivities to normalization granularity and scope. These methods could help clarify whether the observed patterns are specific to CPC or general across representation learning strategies in EEG. Nonetheless, our approach enables clear comparisons and sheds light on normalization-specific behaviors in a widely used SSL framework.

**Real-time applications.** An important practical consideration is the use of recording-level normalization, which assumes access to the entire EEG recording before normalization can be applied. This assumption may not hold in real-time or online applications such as neurofeedback, BCI, or clinical monitoring, where data must be processed incrementally. While such normalization can enhance model stability in offline analyses, its utility is limited when only short segments are available at inference time. Nonetheless, strategies such as collecting short baseline periods at the beginning of a session or during rest blocks may offer a compromise by enabling partial estimation of normalization parameters.

Overall, our findings emphasize that normalization must be tailored to the learning objective. Supervised tasks benefit from aggressive local normalization, while SSL models might require more global, structure-preserving strategies. As EEG research moves toward foundation models and large-scale pretraining, careful consideration of normalization will be essential for robust and generalizable learning.